\documentclass[aps,preprint,byrevtex,showpacs,preprintnumbers]{revtex4}
\newcommand{\be}{\begin{equation}}
\newcommand{\ee}{\end{equation}}
\newcommand{\bea}{\begin{eqnarray}}
\newcommand{\eea}{\end{eqnarray}}
\newcommand{\nn}{\nonumber}
\newcommand{\ep}{\epsilon}

\newcommand{\sxi}{\xi\!\!\!/}
\newcommand{\sth}{\theta\!\!\!/}
\newcommand{\snabla}{\nabla\!\!\!\!/}
\newcommand{\vac}{|0\rangle}
\newcommand{\tvac}{|\tilde{0}\rangle}

\newcommand{\ra}{\rightarrow}
\begin{document}
\preprint{{hep-th/0212017}}
\preprint{AEI-2002-097}
\title{Remarks on type IIB pp waves with Ramond-Ramond fluxes and \\
massive two dimensional nonlinear sigma models}
\author{Nakwoo Kim}
\affiliation{
Max-Planck-Institut f\"ur Gravitationsphysik,
Albert-Einstein-Institut,
Am M\"uhlenberg 1, D-14476 Golm, Germany
}
\email{nakwoo.kim@aei.mpg.de}
\begin{abstract}
We continue the study of supersymmetric type IIB pp-wave solutions
by Maldacena and Maoz (hep-th/0207284), who showed 
Ramond-Ramond five-forms can induce potential terms in the light cone
string actions which are nonlinear sigma models with special holonomy
target spaces. We show that nonvanishing Ramond-Ramond three-forms
provide extra potential terms involving Killing vectors in the string action
and identify the supersymmetry requirements. 
In particular, in solutions with $(1,1)$ worldsheet
supersymmetry, the Killing vectors are required to be self-dual in Spin(7). 
\pacs{11.25.-w, 04.65.+e}
\end{abstract}
\maketitle
\section{Introduction}
Recently Maldacena and Maoz \cite{malmao} have constructed an interesting 
class of supersymmetric pp-wave solutions in ten dimensional type 
IIB supergravity,
which includes the maximally supersymmetric plane-wave solution \cite{blau} 
as a special case.
The nontrivial curvature is supported by null Ramond-Ramond (RR) five-forms
which are non-constant, and it is argued that the light cone
string actions in Green-Schwarz formalism are nonlinear sigma models with
potential terms. When the target space is curved 
the spacetime supersymmetry requires it to have special holonomy. 
For solutions with $(2,2)$ worldsheet supersymmetry
the target spaces are Calabi-Yau four-folds in general, and RR five-forms
give holomorphic superpotential and Killing vector terms, whereas
when one demands only $(1,1)$ supersymmetry the target geometry has 
reduced Spin(7) holonomy most generally and non-constant five-forms
are translated into real harmonic superpotentials on the worldsheet.
These solutions are shown to be exact string backgrounds using the U(4)
formalism \cite{bermal} and also by considering possible higher order
correction terms to the string effective actions \cite{russo}. Thus a large
class of interacting two dimensional field theories, including
integrable models, are added to the list of quantizable RR backgrounds. 
For subsequent works on these pp-wave string theories see 
\cite{cvetic, berkovits, bonelli, hubeny, hikida, gomez, sonn}.

Although the solutions in \cite{malmao} comprise quite general class
of massive nonlinear sigma models, certainly they do not have the 
{\it most general} potential terms. 
The supersymmetry requirements of potential terms in
two dimensional nonlinear sigma models are summarised in \cite{alvarez}.
With ${\cal N}= (2,2)$ the target spaces are K\"ahler and {\it two}
commuting holomorphic Killing vector terms and one holomorphic 
superpotential term can be present. And for ${\cal N}=(1,1)$
the target space can be any real manifold and the potential terms include 
one real superpotential and one Killing vector term. Supersymmetry
also requires that the Lie derivative of the superpotential should be
constant along the Killing vectors. Compared to this, first
of all the field theories from pp-waves are special in the sense that 
the target space is always eight dimensional with special holonomy,
and secondly we see that one Killing vector contribution is missing
both in ${\cal N}=(2,2)$ and ${\cal N}=(1,1)$ solutions. 
It is conceivable that RR three-forms can provide the missing potential
terms on the worldsheet. It is the aim of this paper to show it is indeed
the case. 

In sec. \ref{two} we closely follow and repeat the
analysis of \cite{malmao} with nontrivial RR three-forms 
as well as five-forms. 
We identify the spacetime supersymmetry requirements on the Killing
vectors from RR three-forms and find they are consistent with the
results in \cite{alvarez} as quoted above. 
For the analysis of ${\cal N}=(1,1)$ solutions we find
it more illuminating to assume that the transverse space has
Spin(7) holonomy, and find supersymmetry requires the Killing
vector to preserve the Spin(7) structure. 
In sec. \ref{four} we conclude with a few comments.
\section{IIB PP-wave solutions with Ramond-Ramond backgrounds}
\label{two}
For IIB supergravity in ten dimensions we follow the conventions 
of \cite{schw}, and take the ansatz
\bea
ds^2 &=& -2 dx^+ dx^- + H(x^i) (dx^+)^2 + ds^2_8 ,
\nn \\
\label{ansatz}
F^{(5)} &=& dx^+ \wedge \xi (x^i) ,
 \\
F^{(3)} &=& dx^+ \wedge \theta (x^i) ,
\nn
\eea
where $i=1,2\ldots 8$ denote the transverse eight dimensional
space with Euclidean signature, 
$F^{(5)},F^{(3)}$ are the RR fields and all other fields are set to zero.
The Einstein's equation has only one nontrivial component
\be
\nabla^2  
H = -  \frac{2}{3} \xi_{ijkl}\xi^{ijkl} 
    -  \frac{1}{2} \theta_{ij} \theta^{ij} ,
\ee
and from other field equations and the Bianchi identities 
$\xi$ is anti-self-dual, closed and co-closed,
and $\theta$ is a closed two-form in eight dimensions. 

The supersymmetry transformations for the dilatino $\lambda$
and the gravitino $\psi_a$ are given in terms of a Weyl spinor
$\epsilon$,
\bea
\delta \lambda &=& \frac{1}{24} G_{abc} \Gamma^{abc} \epsilon ,
\\
\delta \psi_a &=& D_a \ep - \Omega_a \ep - \Lambda_a \ep^* ,
\eea
with
\bea
\Omega_a &=& -\frac{i}{480} F^{(5)}_{bcdef} \Gamma^{bcdef} \Gamma_a ,
\nn\\
\Lambda_a &=& -\frac{1}{96} ( \Gamma_a G_{bcd}\Gamma^{bcd} +
2 G_{bcd} \Gamma^{bcd} \Gamma_a ) .
\nn
\eea
$a,b,c \ldots$ are used to denote the ten dimensional frame
indices and the gamma matrices are constants subsequently. $G_{abc}$
is the complexified three-form, and since we have set Neveu-Schwarz (NS) 
fields to zero $G$ is pure imaginary, i.e. $G = i F^{(3)}$.

Given the ansatz Eq.(\ref{ansatz}), it is natural to employ the following
decomposition of $\ep$, 
\bea
\ep &=& -\frac{1}{2} \Gamma_+ \Gamma_- \ep
      -\frac{1}{2} \Gamma_- \Gamma_+ \ep
\equiv \epsilon_+ + \epsilon_- .
\nn
\eea
It is important to note that $\ep_+,\ep_-$ have opposite chiralities
in SO(8). Now we write down the Killing spinor equations in terms
of $\ep_\pm$. First by setting the variation of the dilatino to zero,
\be
\label{dila}
\sth \ep_+ = 0 ,
\ee
and from $\delta\psi_a=0$ we get
\bea
\label{m-pm}
\partial_- \ep_+ = \partial_- \ep_- = 0 ,
\\
\label{p-p}
\partial_+ \ep_+ + \frac{i}{8} \sth \ep^*_+ = 0 ,
\\
\label{t-p}
\nabla_i \ep_+ = 0 ,
\\
\label{p-m}
\partial_+ \ep_- +
\frac{i}{2} \sxi \ep_- +
\frac{i}{4} \sth \ep^*_- -
\frac{1}{4} \Gamma_- \snabla H \ep_+ = 0 ,
\\
\label{t-m}
\nabla_i \ep_- -
\frac{i}{4} \Gamma_- \sxi \Gamma_\mu \ep_+ +
\frac{i}{16} \Gamma_- ( \Gamma_\mu \sth - 2 \sth \Gamma_\mu ) \ep^*_+
= 0 ,
\eea
where $\sxi = \frac{1}{4!} \xi_{ijkl} \Gamma^{ijkl}, \sth = \frac{1}{2} 
\theta_{ij} \Gamma^{ij}$. First from Eq.(\ref{m-pm}) we see that
the Killing spinors are independent of $x^-$.

We find it useful to recall here the interpretation of different
Killing spinor solutions given in \cite{malmao}.
The Killing spinor solutions with nonvanishing $\ep_+$ are
related to dynamical supersymmetries in the light cone worldsheet action
when it is rearranged to give a nonlinear sigma model. 
On the other hand for the other
half of Killing spinor components $\ep_+$ can be set to zero
and we can try to solve the remaining equations of $\ep_-$ . 
These Killing
spinors are related to the kinematic part of the supersymmetries
in the light cone gauge, and in the nonlinear sigma models
they are related to the number of free fields. Note that 
these Killing spinors in general depend on $x^+$, and the
$x^+$ dependence of $\ep_+$-nonvanishing Killing spinors
can be ignored by exploiting that we are free to superpose two 
different types of Killing spinor solutions. 
Since our aim here is to show that RR three-forms give potentials involving 
Killing vector terms to the string worldsheet action, and in flat spaces
the isometries are trivial, we concern ourselves particularly with 
nontrivial target geometry. We thus look for Killing spinors 
with nonzero $\ep_+$ only.

\subsection{Solutions with four supercharges}\label{22}
In order to have two linearly independent complex spinors with
nonvanishing $\ep_+$, we see first from Eq.(\ref{t-p}) that the transverse
eight-dimensional space should most generally be a Calabi-Yau four-fold.
It is thus useful to consider how $\xi,\theta$, originally
in ${\bf 35}, {\bf 28}$ of SO(8) are decomposed into SU(4). 
\bea
{\bf 35} &\ra & {\bf 15} + {\bf 10} + {\bf \overline{10}}
,\\
{\bf 28} &\ra & {\bf 15} + {\bf 6} + {\bf \overline{6}} + {\bf 1} .
\eea
When we choose the basis where the Killing spinors
satisfy
\be
\Gamma_{12}\ep_+=\Gamma_{34}\ep_+=\Gamma_{56}\ep_+=\Gamma_{78}\ep_+ , 
\ee
it becomes natural to choose the Fock space notations with
$\gamma_{\mu} = \Gamma_{2\mu-1} + i \Gamma_{2\mu}, \mu=1,2,3,4$ 
and states $\vac,\tvac$ satisfying 
$\gamma^{\mu}\vac=\gamma^{\bar{\mu}}\tvac=0$.

Now $\ep_+$ can be written as
\be
\label{ep22}
\ep_+ = \alpha \vac + \zeta \tvac ,
\ee
where $\alpha,\zeta$ are complex constants, and we used 
the fact that in Calabi-Yau spaces a gauge choice for the spin 
connections can be made such that the Killing spinors are constants.
Using this basis, it is easy to see from Eq.(\ref{dila}) that
only ${\bf 15}$, a traceless $(1,1)$-form, can be nonzero since they 
annihilate the Killing spinor $\ep_+$ in Eq.(\ref{ep22}) for arbitrary
$\alpha,\zeta$. When ${\bf 6}$ and ${\bf \bar{6}}$, holomorphic and 
anti-holomorphic two-forms of SU(4) respectively, are dual to each
other with respect to the holomorphic four-form 
$\epsilon_{\mu\nu\lambda\rho}$, Eq.(\ref{ep22}) can be satisfied with a 
relation between $\alpha,\zeta$, making the number of worldsheet 
supercharges reduced to two, i.e. ${\cal N} = (1,1)$. Together 
with ${\bf 15}$, there are 21 components of $\theta$ 
which now satisfy Eq.(\ref{ep22}) and it is better described 
as ${\bf 21}$ of Spin(7) in SO(8). This
will be studied in more detail in the next subsection and here 
we consider the solutions with arbitrary $\alpha,\zeta$.

Using the fact that $\ep_+,\ep_-$ have opposite SO(8) chiralities
we can write
\bea
\ep_- &=& \Gamma_- ( \beta_{\bar{\mu}} \gamma^{\bar{\mu}}\vac +
\delta_\mu \gamma^\mu \tvac ) .
\eea
Now it is straightforward to find equations for 
$\alpha,\zeta,\beta_{\bar{\mu}},\delta_\mu$ from the Killing spinor
equations. Following \cite{malmao} we introduce a holomorphic tensor
$\xi_{\mu\nu}\equiv\frac{1}{3!} \xi_{\mu\overline{\lambda\rho\sigma}}
\ep^{\overline{\lambda\rho\sigma\nu}} g_{\overline{\nu}\nu}$
and a hermitian tensor, 
$\xi_{\mu\overline{\nu}} \equiv \frac{1}{2} g^{\lambda\overline{\lambda}} 
\xi_{\mu\overline{\nu}\lambda\overline{\lambda}}$. 

From Eq.(\ref{p-m}), we have 
\bea
2i (-\beta_{\bar{\mu}} \xi^{\bar{\mu}}_{\;\;\bar{\nu}} +
\delta_{\mu} \xi^{\mu}_{\;\;\bar{\nu}} )
-\frac{i}{2} \theta^{\bar{\mu}}_{\;\;\bar{\nu}} \delta_{\bar{\mu}}
&=& \frac{\alpha}{4} \partial_{\bar{\nu}} H
\label{hh1} ,
\\
2i (+\beta_{\bar{\mu}} \xi^{\bar{\mu}}_{\;\;\nu} -
\delta_{\mu} \xi^{\mu}_{\;\;\nu} )
+\frac{i}{2} \theta^{\mu}_{\;\;\nu} \beta_{\mu}
&=& \frac{\zeta}{4} \partial_{\nu} H ,
\label{hh2}
\eea
with $\beta_\mu\equiv\beta^*_{\bar{\mu}},
\delta_{\bar{\mu}}\equiv\delta^*_\mu$.
Eq.(\ref{t-m}) gives
\bea
\nabla_{\bar{\mu}} \beta_{\bar{\nu}} - 
i\zeta \xi_{\bar{\mu}\bar{\nu}} &=& 0 ,
\\
\nabla_\mu \delta_\nu - i\alpha \xi_{\mu\nu} &=& 0 ,
\\
\nabla_\mu \beta_{\bar{\nu}} + i\alpha \xi_{\mu\bar{\nu}} +
\frac{i}{4} \zeta^* \theta_{\mu\bar{\nu}} &=& 0 ,
\\
\nabla_{\bar{\mu}} \delta_{\nu} + i\zeta \xi_{\bar{\mu}\nu} -
\frac{i}{4} \alpha^* \theta_{\mu\bar{\nu}} &=& 0 .
\eea

Of course apart from the terms involving $\theta$'s the above
equations are the same as the ones found in \cite{malmao}, and
we can proceed in the same spirit to find the solutions for
$\beta_{\overline{\mu}},\delta_\mu$ in terms of arbitrary $\alpha,\zeta$
and identify the requirements on $\xi,\theta$ from consistency. 

By exploiting the properties
like closure and anti-self-duality of $\xi,\theta$ and considering
the integrability conditions we find we can write
\bea
\xi_{\mu\nu} &=& 2 \nabla_\mu \nabla_\nu W ,
\\
\xi_{\mu\overline{\nu}} &=& 2 \nabla_\mu \nabla_{\overline{\nu}} G_1 ,
\\
\theta_{\mu\overline{\nu}} &=& 8 \nabla_\mu \nabla_{\overline{\nu}} G_2 ,
\eea
with the coefficients chosen for later convenience.
$W$ is a holomorphic function and $G_1,G_2$ are real harmonic functions
which serve as the potentials for holomorphic Killing vectors
$V^1_{\mu} = i \nabla_\mu G_1$ and $V^2_{\mu} = i \nabla_\mu G_2$. The
fact that $G_1,G_2$ are harmonic is obvious from the group theory
consideration that they are ${\bf 15}$, i.e. $(1,1)$-form which is
{\it traceless}.

The solution for $\ep_-$ is given by
\bea
\beta_{\overline{\mu}} &=& 
2i ( - \alpha 
\nabla_{\overline{\mu}} G_1
+ \zeta 
\nabla_{\overline{\mu}} 
\overline{W}
+ \zeta^* 
\nabla_{\overline{\mu}} 
G_2 ) ,
\\
\delta_{\mu} &=& 
2i ( - \zeta 
\nabla_{\mu} G_1
+ \alpha
\nabla_{\mu} 
W
+ \alpha^* 
\nabla_{\mu} 
G_2 ) .
\eea

When they are substituted into Eqs.(\ref{hh1}) and (\ref{hh2})
we find from consistency
\bea
\nabla_\nu G_2 \nabla^\nu \nabla_\mu G_1 -
\nabla_\nu G_1 \nabla^\nu \nabla_\mu G_2   
&=& 0  ,
\nn\\
\nabla_{\nu} ( \nabla^{\nu} G_1 \nabla_\mu W ) &=& 0  ,
\nn\\
\nabla_{\nu} ( \nabla^{\nu} G_2 \nabla_\mu W ) &=& 0 ,
\nn
\eea
and the complex conjugates. It is obvious what they mean. As 
promised, the spacetime supersymmetry requires that the two
Killing vectors should commute with each other and the Lie
derivative of the holomorphic superpotential along the Killing
vectors should vanish. This precisely matches the field
theory supersymmetry conditions in \cite{alvarez}.

Finally the metric is given in terms of
$$
H = -32 g^{\mu\overline{\nu}} (
\nabla_\mu G_1 {\nabla}_{\overline{\nu}} G_1 +
\nabla_\mu G_2 {\nabla}_{\overline{\nu}} G_2 +
\nabla_\mu W \nabla_{\overline{\nu}} \overline{W}
).
$$
\subsection{Solutions with two supercharges}\label{11}
The existence of one Killing spinor solution to $\nabla_i \ep_+=0$
implies that the transverse eight-dimensional space has a reduced
holonomy Spin(7) in general. The Killing spinor which is left invariant
under Spin(7) subgroup of SO(8) will be denoted as $\eta$ and is chosen
to be real. As it is well-known Spin(7) holonomy is 
characterised by the Cayley four-form $\Psi$
which can be constructed from the Killing spinor
\be
\Psi_{ijkl} = \eta^T  \Gamma_{ijkl} \eta ,
\ee
which is covariantly constant by construction and self-dual in SO(8)
when we take the convention that $\xi$ is anti-self-dual. We assume
that $\eta$ is normalised appropriately and $\Psi$ can take values
$0,\pm 1$. In the standard basis the non-zero components of $\Psi$ are 
given as
\bea
1 &=& \Psi_{1234}  =
\Psi_{5678}  =
\Psi_{3478}  =
\Psi_{2468}  =
\Psi_{2367}  
\nn\\
&= & \Psi_{1368} = 
\Psi_{1256} =
\Psi_{1357} =
\Psi_{1458} =
\Psi_{2457} 
\nn\\
&= & \Psi_{1287} = 
\Psi_{1476} =
\Psi_{3465}  =
\Psi_{2385} . 
\nn
\eea
The basic identity involving $\Psi$ is 
\be
\label{basic}
\Psi_{ijkp} \Psi^{lmnp} = 
\frac{1}{6} \delta^{[l}_{[i} \delta^{m}_{j} \delta^{n]}_{k]}
- \frac{1}{4} \Psi_{[ij}^{\;\;\;\;[lm} \delta_{k]}^{n]} ,
\ee
which will prove useful in verifying the statements in the following 
discussions. 

Now in order
to solve the Killing spinor equations involving $\xi,\theta$, we first
recall that
\bea
\label{35}
{\bf 35}_{asd} &\ra & {\bf 35}  ,
\\
\label{28}
{\bf 28} \quad\; &  \ra & {\bf 21} + {\bf 7} .
\eea
It turns out that ${\bf 35}$ of Spin(7) can be alternatively described as a
traceless symmetric rank-two tensor when we make use of $\Psi$. It is
straightforward to show that
\be
(\Psi \cdot \xi)_{ij} \equiv \Psi_{iklm} \xi_{j}^{\;\;klm}
\ee
is symmetric and traceless. 

The decomposition of the adjoint 
representation Eq.(\ref{28}) and its implication on the solution of 
Eq.(\ref{dila}) is rather
famous. Projection operators for two-forms in eight dimensions can be 
explicitly constructed using $\Psi$, so that
\be
\label{sd}
\theta^{\pm}_{ij} =
\lambda_{\pm}
\Psi_{ij}^{\;\;\;\;kl}
\theta^{\pm}_{kl} ,
\ee
where $\theta^+$ is ${\bf 21}$ with $\lambda^+ = \frac{1}{2}$, while
$\theta^-$ is ${\bf 7}$ with $\lambda^- = -\frac{1}{6}$. This was
first considered in \cite{corrigan} as a generalization of
the four-dimensional self-dual gauge fields, and extended to nontrivial
special holonomy manifolds in \cite{acharya}. It is customary to
call {\bf 21} self-dual and {\bf 7} anti-self-dual.
Since $\ep_+$ is
proportional to $\eta$, the invariance of the dilatino means that
we should keep ${\bf 21}$ while ${\bf 7}$ should be set to zero.
This is consistent with what we observed in the previous subsection with
transverse Calabi-Yaus. When we demand two complex $\ep_+$ spinors we
keep only ${\bf 15}$ of SU(4), but when ${\bf 6}$ and ${\bf \bar{6}}$
are nonzero and dual to each other we still have one spinor solution
to Eq.(\ref{dila}) $\eta=\vac+\tvac$.

Now we take the following ansatz for Killing spinors,
\bea
\ep_+ = \alpha \eta , 
\quad\quad\quad
\ep_- = -i \Gamma_- v_i \Gamma^i \eta ,
\eea
where $\alpha$ is a constant and $v_i$ is an unknown nonconstant
vector to be determined. From Eq.(\ref{t-m}) we have
\be
\label{7-1}
\nabla_i v_j = \frac{\alpha}{4!} ( \Psi \cdot \xi )_{ij} +
\frac{1}{8} \theta_{ij} \alpha^* .
\ee
In deriving this and other equations we have chosen the gauge 
for spin connections which makes $\eta$ constant, like the 
discussions about Calabi-Yau four-folds in the last subsection. 
A proof that it is possible also with exceptional holonomy manifolds 
can be found for instance in \cite{bilal}.

Since $\Psi,\xi$ are closed, $( \Psi \cdot \xi )_{ij} $ is curl-free.
Therefore we can write locally
\be
\frac{1}{4!} (\Psi\cdot\xi)_{ij} = \nabla_i\nabla_j U ,
\ee
where $U$ is a real harmonic function since $(\Psi\cdot\xi)_{ij}$ is 
traceless, and $U$ becomes the real superpotential of ${\cal N}=(1,1)$
nonlinear sigma models in the light cone lagrangian.
Then Eq.(\ref{7-1}) implies that the gauge potential for $\theta$
can be chosen to be a Killing vector, i.e.
\be
\frac{1}{8} \theta_{ij} = D_i G_j ,
\quad\quad\quad
D_{(i} G_{j)} = 0 .
\ee
When we substitute
\be
v_i = \alpha \nabla_i U + \alpha^* G_i ,
\ee
into Eq.(\ref{p-m}) we get a consistency 
condition on the Killing vector $G$,
\be
D_i ( G^j \nabla_j U) = 0
\quad\quad
{\rm or}
\quad\quad
{\cal L}_G U = {\rm const}
\ee
which matches with the condition for potential terms of
${\cal N}=(1,1)$ supersymmetric nonlinear sigma models. 
When we integrate what remains we obtain
\be
H = -4 ( G_i G^i + \nabla_i U \nabla^i U ) ,
\ee
which serves as the scalar potential of the light cone worldsheet
lagrangian.

Before we finish let us point out that an alternative 
interpretation can be given to the requirement that the Killing
vector should be in the adjoint representation ${\bf 21}$ of Spin(7).
We note that for any Killing vector $K$,
\bea
{\cal L}_K \Psi &\equiv& ( d i_K + i_K d ) \Psi
\nn\\
&=& \nabla_j K^n \; \Psi_{nklm} 
\;\;
dx^j \wedge dx^k \wedge dx^l \wedge dx^m
\nn\\
&=& ( \nabla_j K^n )^- \Psi_{nklm}
\;\;
dx^j \wedge dx^k \wedge dx^l \wedge dx^m .
\nn
\eea
In order to get the third line Eq.(\ref{basic}) and Eq.(\ref{sd}) are used. 
We thus see that the Killing vector $G$'s being ${\bf 21}$ means it 
preserves the Spin(7) structure, in the sense that the Lie
derivative of $\Psi$ vanishes along $G$.
\section{Discussions}
\label{four}
In this paper we have presented a general class of IIB pp-wave solutions 
with Ramond-Ramond backgrounds. In the light cone gauge the bosonic string
action can be simply read off from the metric and it is obvious we
have nonlinear sigma models with eight dimensional special holonomy
manifold target spaces followed by potential terms given by $H$. 
Then the worldsheet supersymmetries inherited by the Killing spinors
found above dictates how the terms with fermions should be written.
In particular the non-vanishing RR fields give fermionic mass terms
or Yukawa couplings more generally. 

The potential terms of supersymmetric nonlinear sigma models 
are studied in \cite{alvarez}. The analysis does not make
use of the superfield formalism; it started with the most general
lagrangian and supersymmetry transformation rules allowed by
Lorentz invariance and found the consistency conditions.
The potential is given in terms of 
one holomorphic superpotential and two commuting holomorphic
Killing vectors for ${\cal N}=(2,2)$ models and one real superpotential
and one Killing vector for ${\cal N}=(1,1)$ solutions. 
The Lie derivatives of the superpotentials along the Killing vectors
are required to be constants. 
The supergravity analysis presented in \cite{malmao} and here is found 
to be consistent with the field theory results,
but in general the spacetime supersymmetry is more restrictive. 
It is particularly distinctive with ${\cal N}=(1,1)$ solutions;
the superpotential is a harmonic function and the Killing vector 
should be self-dual with respect to Spin(7). For ${\cal N}=(2,2)$
solutions the Killing potentials are required to be harmonic.

Perhaps it is useful to consider an example of the pp-wave with
nonzero three-form. In flat transverse space mass terms can be given
to the worldsheet fields by a holomorphic Killing vector, for instance
one can consider a pure RR three-form background such as 
\bea
F^{(3)}_{+12} = 
F^{(3)}_{+34} = 
F^{(3)}_{+56} = 
F^{(3)}_{+78} = 
m .
\label{s28}
\eea
This solution is already considered in \cite{bena} where generic
supersymmetric plane-wave solutions with nonvanishing RR fields are
studied. This solution in fact preserves 28 spacetime supersymmetries.
What is special with this solution is that the worldsheet
fields all have the same mass. With bosons it is obvious from the metric
and the worldsheet supersymmetry guarantees that fermion masses are
the same. In other words, the light cone string spectrum
of this background is the same as that of the maximally supersymmetric 
solution with nontrivial five-form \cite{metsaev}. This of course is a natural
consequence of our claim that for ${\cal N}=(2,2)$ solutions the same 
Killing vector terms on the worldsheet can come from either RR five-forms 
or three-forms.
Our result is also consistent with the observation made in \cite{bena}
that the plane-wave solutions can be superposed. In this paper we have
extended it to nontrivial target geometries and non-constant form fields,
i.e. general pp-waves. 
Although we expect the degeneracy will be lifted once string interactions
are taken into account, it will be interesting if we can find a field theory
dual of this solution using the Penrose limit as in \cite{bmn, bfp}. It 
is not immediately obvious how or whether the solution Eq.(\ref{s28}) 
can be obtained from an AdS solution as the Penrose limit \cite{bena}. 
It might be also worth mentioning that our
results can be used to regularize the orbifolds of solutions found
in \cite{bena} by replacing the orbifolded part of the target spaces 
with, e.g., Eguchi-Hanson space.

In \cite{russo} pp-wave solutions with nonvanishing 
NS and RR three-forms are considered and the authors concluded
that three-forms cannot induce worldsheet interactions without
breaking supersymmetry. Our result does not contradict theirs,
since in flat target spaces the Killing vectors can give mass
terms at most. It is essentially the target space curvature and 
the superpotential which make the worldsheet action interacting,
but the message of our analysis is that the RR three-forms can
be succinctly incorporated into interacting models.

Finally we reckon it is an important enterprise to study the field theoretical
properties of the massive nonlinear sigma models found in this paper.
They are special in the sense that although manifestly non-conformal 
they can be embedded into exact superconformal theories. By working
out the quantum corrections one might be able to identify the hidden
conformal covariance of these massive two dimensional field theories.
It should also help rectifying the definition of supersymmetric D-branes 
in the class of nonlinear sigma models considered in this paper.

\acknowledgments
We wish to thank S. Fredenhagen, V. Schomerus, and
S. Theisen for helpful discussions at various stages of this work,
B. Acharya for correspondences, and especially P. Kaste for
drawing our attention to \cite{bilal}. The conversation with P.
Kaste took place during the 35th International Symposium Ahrenshoop
held at Berlin, 26-30 August 2002. We thank the organisers for hospitality.
This work was supported by the DFG - German Science Foundation.

\end{document}